\documentclass[manuscript]{acmart}
\usepackage{todonotes}
\usepackage{wrapfig}

\copyrightyear{2022}
\acmYear{2022}
\setcopyright{rightsretained}
\acmConference[RecSys '22]{Sixteenth ACM Conference on Recommender Systems}{September 18--23, 2022}{Seattle, WA, USA}
\acmBooktitle{Sixteenth ACM Conference on Recommender Systems (RecSys '22), September 18--23, 2022, Seattle, WA, USA}
\acmDOI{10.1145/3523227.3551480}
\acmISBN{978-1-4503-9278-5/22/09}

\begin{document}

\title{Recommender Systems and Algorithmic Hate}

\author{Jessie J. Smith}
\email{jessie.smith-1@colorado.edu}
\orcid{}
\author{Lucia Jayne}
\email{lucia.jayne@colorado.edu}
\author{Robin Burke}
\email{robin.burke@colorado.edu}
\affiliation{%
  \institution{University of Colorado, Boulder}
  \city{Boulder}
  \state{Colorado}
  \country{USA}
  \postcode{80027}
}

\renewcommand{\shortauthors}{Smith et al.}

\begin{abstract}
  Despite increasing reliance on personalization in digital platforms, many algorithms that curate content or information for users have been met with resistance. When users feel dissatisfied or harmed by recommendations, this can lead users to hate, or feel negatively towards these personalized systems. Algorithmic hate detrimentally impacts both users and the system, and can result in various forms of algorithmic harm, or in extreme cases can lead to public protests against ``the algorithm'' in question. In this work, we summarize some of the most common causes of algorithmic hate and their negative consequences through various case studies of personalized recommender systems. We explore promising future directions for the RecSys research community that could help alleviate algorithmic hate and improve the relationship between recommender systems and their users.
\end{abstract}



\maketitle

\section{Introduction}
Personalized recommender systems have transformed the way we consume information in the digital world. Personalization algorithms use historical user data to predict what content or information the user would like to be exposed to in the future. On social media platforms where users are exposed to large amounts of content, personalizing content `feeds' is an essential component for combating information overload. In 2009, Facebook recognized personalization as an essential feature to implement on their platform to encourage users to continue `friending' people without feeling over-saturated with information from posts made by their growing friends list. Once Facebook made the decision to algorithmically curate user's timelines, this set a precedent for other social media platforms that expose large amounts of content to their users. Over time, this trend towards personalization has prompted increasing backlash from users, and in some cases has led to public protests against ``the algorithm.''


Though previous work has summarized different characteristics of algorithms that might provoke negative feelings from users \cite{jussupow2020we}, little work to date has focused specifically on algorithmic hate towards recommendation systems and the negative impacts it can have on the relationship between user and system. In this work, we seek to explore this gap. For the scope of this paper we define \textbf{algorithms} to be recommendation or ranking systems, specifically those that attempt to \emph{personalize} content based on its perceived relevance for users. We define \textbf{hate} as a feeling of aversion, dislike, or irritation \cite{ytre2021folk}; we use these four words interchangeably throughout this paper. We define \textbf{algorithmic hate} as a version of the definition provided by \cite{jussupow2020we}: \emph{``biased assessment of an algorithm which manifests in negative behaviours and attitudes towards the algorithm.''} Algorithmic hate in recommendation has a wide range of causes and effects, which we explore in this formative work. Specifically, in this paper we seek to understand how personalized recommendation and ranking algorithms have led users to react with algorithmic hate, and the negative consequences that can result from that experience. Using this knowledge, we outline research directions that can improve the relationship between the user and the algorithm -- with focus on how the RecSys research community can help prevent algorithmic hate in the future.

\subsection{Case-Studies of Algorithmic Hate in Recommendation}
To begin, we describe some of the most famous accounts of algorithmic hate, and features of personalized recommendations that are often associated with negative feelings towards ``the algorithm.'' Starting with Twitter, in 2016 the widely used \#RIPTwitter hashtag was a reflection of users' dislike towards Twitter's decision to switch from their reverse chronological timeline to a timeline based on personalized recommendations \cite{devito2017algorithms}. In 2017, the hashtag \#RIPInstagram emerged in response to Instagram's decision to implement algorithmic personalization on their platform as well. Emotional disdain for personalized recommendations on social media is not unique to Twitter or Instagram. A previous study showed that when any digital platform modifies their ``organic'' newsfeeds or timelines to a personalized system, user responses are \emph{``self-motivated and highly emotional,''}  \cite{mahnke2018please}. This emotional response stems from a strong connection that users make with their information feed. \emph{``Algorithmic personalization has turned abstract information feeds into personal and intimate information spaces, which prompt forms of ``connective action'' when modified without consent. Feelings of powerlessness and loss of control especially trigger such responses,''} \cite{mahnke2018please}. In another study, researchers came to the same conclusion, describing how users' emotional response against Instagram's personalized feed was a result of a loss of agency and autonomy on the platform \cite{skrubbeltrang2017ripinstagram}. We explore this theme further in Section \ref{agency}.

\begin{wrapfigure}{r}{0.5\textwidth}
  \begin{center}
    \frame{\includegraphics[width=0.5\textwidth]{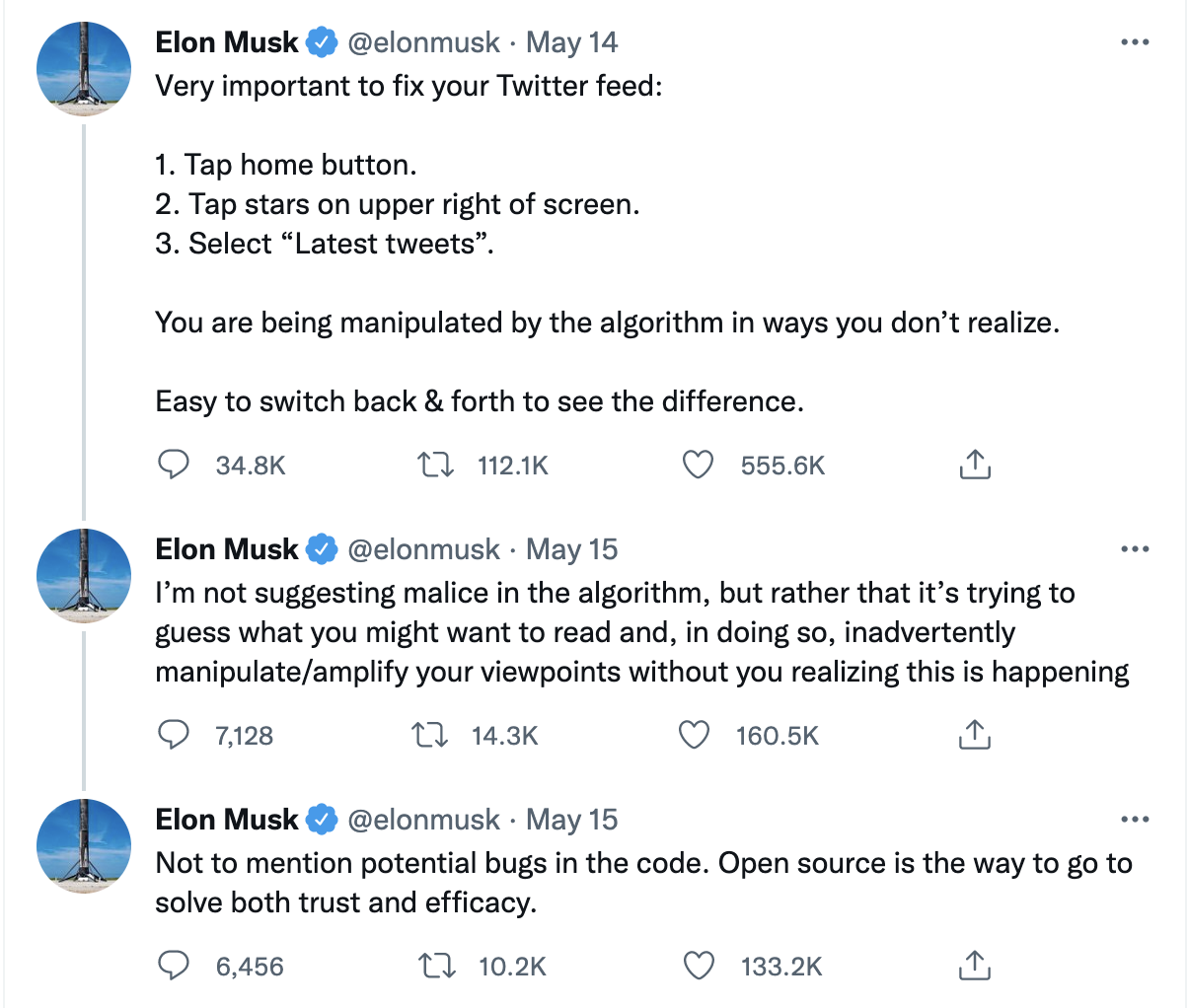}}
  \end{center}
  \caption{Tweets from Elon Musk About Twitter's Personalized Feed.}
  \label{fig:Elon-Musk}
\end{wrapfigure}

In more recent news, after Elon Musk's attempt to purchase Twitter in April 2022, Musk publicly invoked algorithmic hate towards Twitter's home feed, expressing that \emph{``you are being manipulated by the algorithm in ways you don't realize,''} as shown in Figure \ref{fig:Elon-Musk}. Whether a recommendation platform intends to manipulate users or not, just the \emph{feeling} of being manipulated can cause users to feel algorithmic anxiety \cite{alg-anxiety}.

Lowered user experience is another common feature associated with algorithmic hate in personalization. In the news environment, one study showed that algorithmically curated timelines increased users' exposure to junk news by 3\% \cite{twitter2021comparison}. Algorithmic hate has also been associated with algorithms that break the law -- whether the recommendation results circumvented regulation intentionally or not. For instance, Facebook has been under scrutiny for using its recommendation systems in a discriminatory manner. In June 2022, the company settled a lawsuit with the US Department of Justice because their display of housing ads based on ``Lookalike Audience’’ violated the Fair Housing Act as it allowed for discrimination on the basis of race, sex and other factors \cite{guardian2022}. From a political standpoint, algorithmic hate can be linked to the perception of algorithmic bias by different political parties. The Hostile Media Effect (HME) is described as the perception that neutral news coverage is biased. This effect has been observed across all political parties. One study observed that HME is particularly strong for members of the Republican party in the US, where political leaders have previously amplified false narratives that they are being unfairly censored by algorithms online \cite{polarization_2021}. Additionally, HME could be enhanced by the lack of transparency in personalized user feeds. In one study, participants struggled to differentiate personalized news from targeted commercial advertisements \cite{user_news_2020}. This kind of uncertainty around personalized information feeds can make it challenging to develop AI regulations that satisfy users from different political parties. For example, in the US,  Republican legislators tend to advocate for reducing algorithmic censorship, while US Democrats have active efforts to prohibit discrimination by algorithms based on race, sex, and other characteristics -- which can increase online censorship \cite{dem_2021}. This political divide between users' values adds to already-existing tension and hate against personalized algorithms.

Another feature of personalization that is a target for algorithmic hate is the prevalence of echo chambers or feedback loops. Studies have shown that echo chambers can arise in the digital world regardless of the influence of algorithms. In one study, researchers modeled how a digital environment that is not polarized in the beginning will become polarized over time, due to the mechanics of social influence in digital spaces \cite{ecochamber2020}. In these cases, although algorithms might only be enhancing existing differences, they are often blamed entirely for the polarization that users experience. Conversely, some studies have shown that echo chambers \emph{can} arise purely from personalization. In fact, echo chambers and feedback loops are sometimes considered a key feature that is necessary for the success of a recommender system. For example, in a video recommendation platform such as YouTube, finding yourself in an echo chamber of videos that you like seeing could be an indication of very accurate recommendations \cite{pariser2011filter}. However, when echo chambers that arise from personalized algorithms cause users to expose themselves to detrimental content that they wouldn't have otherwise (e.g., conspiracy theories), these are considered degenerate feedback loops \cite{jiang2019degenerate, youtubeconspiracy}. Whether feedback loops are a feature or a bug in a given recommender system, when users become aware that they are in one, criticism and negativity towards the algorithm often follow.


\section{What Causes Algorithmic Hate?}
In an effort to explore the negative consequences of algorithmic hate, we now examine what leads users to feel negatively towards personalization. In this section, we explore the various design decisions in recommender systems that have left users feeling silenced, dissatisfied, or harmed on various platforms -- and how each case study negatively impacted relationships between user and system.
\label{causes}

\subsection{Lack of Knowledge}
Recommender systems used in online platforms are highly complex, created from a combination of algorithmic elements including feature extraction, ranking, flagging, re-ranking, scoring, and/or clustering using a variety of technologies and various forms of machine learning. Recommendation systems built from interactions between large, complex sub-systems can be difficult to understand even for those who create them. For ordinary users, this is even more of a challenge. The term \emph{algorithmic skill} is defined as an understanding of how algorithms impact visibility of content online \cite{klawitter2018s}. \citet{gruber2021algorithm} categorize algorithmic awareness as a meta-skill because it directly impacts other necessary digital skills. Within algorithmic awareness, users often fall into the sceptical/critical category or unaware category \cite{2019society}. We can think of variations in algorithmic skill as an aspect of the oft-cited \textit{digital divide}, which can be segmented into three aspects: (1) access to internet and equipment; (2) skills (e.g., algorithmic awareness); and (3) general benefits for users. The digital divide reflects existing inequality in society and is exacerbated by educational inequalities and differential opportunities for developing internet skills  \cite{algorithm_divide_2021}. Taking this into consideration, one potential contributor to algorithmic hate is the lack of knowledge that users have about these complex systems, evidenced by the monolithic term ``the algorithm'', for what are in fact collections of algorithms, policies and procedures.

Previous work has also explored how increasing user understanding in recommendations can improve user trust in the platform, and conversely has shown how lack of knowledge can fuel lack of trust towards the platform that hosts the system \cite{komiak2006effects}. In extreme cases, mistrust can lead users to believe that they are being manipulated by the algorithm -- whether it is intentional or not -- as referenced in Figure \ref{fig:Elon-Musk}. Trust is closely tied to the level of control that users feel they have online. One study showed that even giving users minimal control over their recommendations positively impacted their desire to use a recommendation algorithm \cite{control_2018}. We explore this theme further in Section \ref{agency}.

When users aren't aware of how personalized algorithms work, they create their own theories about how they work -- known as \emph{folk theories} \cite{eslami2016first}. Related to folk theories are \emph{algorithmic imaginaries} which are defined as \emph{``the way in which people imagine, perceive and experience algorithms and what these imaginations make possible,''} \cite{bucher2017algorithmic}. When users generate their own theories or imaginaries about a platform, they might be driven to believe that certain actions on the platform will yield specific results from the algorithm (e.g., `commenting on videos sends more of a positive signal to the algorithm than liking videos'). With respect to folk theories, the actions that users take on personalized platforms are often to achieve specific goals or objectives they have. However, when users' theories about how the algorithm works are incorrect, and their actions do not yield the results they would have liked (thus preventing them from achieving their goals) -- this can lead to algorithmic irritation \cite{ytre2021folk}. One study surfaced this irritation by exploring users' \emph{``Theory of the Recommender,''} and described how users with incorrect theories might take unconventional actions to unsuccessfully `control' their recommendations \cite{ghori2022does}.

\subsection{Algorithmic Harm}
Another cause of algorithmic hate is algorithmic harm -- adverse effects experienced by any stakeholders of an algorithmic system. \emph{Harm} can be defined in many ways, two common definitions include harms of allocation and representation. In recommendation, harms of allocation affect the opportunities that are received by stakeholders of the system, while harms of representation affect the way we perceive or represent the world \cite{ekstrand2021fairness, crawford2017trouble, suresh2021framework, barocas2017problem}.

One example of representational harm is algorithmic exclusion, where \emph{``algorithms construct and reconstruct exclusionary structures within a bounded sociotechnical system, or more broadly across societal structures,''} \cite{simpson2021you}. In one study on the short video recommendation platform, TikTok, some LGBTQIA+ creators experienced exclusion of their identity and/or stereotypical depictions of their identity, reporting that they felt \emph{``silenced''} by the recommender system's content reporting tool \cite{simpson2021you}.

TikTok also previously received hate from LGBTQIA+ content creators who reported that their content was being suppressed by the recommender system. In 2020, the platform issued an apology for shadow-banning (limiting the discovery of content without telling users) certain LGBTQIA+ content and hashtags \cite{ryan2020tiktok}. This was an example of allocative harm leading to algorithmic hate. However, under-representation is not the only allocative harm experienced in recommendation, sometimes over-representation can be just as harmful. For example, when recommendation algorithms make LGBTQIA+ content highly discoverable to the wrong audience, it can lead to targeted digital attacks towards those content creators \cite{devito2021values}.

Popularity bias in recommendations (when popular items are recommended frequently and less popular items are recommended rarely) is another example of allocative harm -- the providers of less-popular items are not given the same opportunities to be discovered as the providers of more popular items. On a music recommendation platform like Spotify, popularity bias for musicians (providers of recommended items) could lead to the allocative harm of under-representation for new artists on the platform \cite{ferraro2019music}. However, representative and allocative harm do not always lead to algorithmic hate -- if users aren't aware that they are being harmed by the system, they will not experience dislike for the system due to this harm. In order for users to recognize that they are being harmed by the system, they must first gain a basic understanding of how the system is functioning, which as we explained in the previous section is not necessarily common knowledge.

\subsection{Objective Misalignment Between User(s) and System}
\label{misalignment}
Originally, personalized recommendation sought to align user goals and algorithmic objectives. However, in more complex systems this alignment might not be so straight-forward. In many personalized systems, recommendations are designed to cater to the needs of multiple stakeholders \cite{abdollahpouri2017recommender}. Consider again the example of Spotify, where one group of stakeholders provides content for recommendation (e.g., podcasters), while another group of stakeholders consumes the recommended content. These stakeholder groups have competing goals (and diverse interests within each group) that the system must trade off against each other, meaning that it cannot be fully aligned with the goals of any particular stakeholder group \cite{burke2017multisided}. In the case of Facebook, the platform caters to many stakeholders -- including end-users, advertisers, content creators, and moderators. In previous years, Facebook's objective to serve personalized advertisements was an attempt to align objectives with both advertisers and end-users. However, previous research uncovered that end-users on Facebook often feel negatively towards targeted advertisements, causing them to feel irritated and skeptical of the algorithm, and in some cases resulted in users opting out of targeted ads altogether \cite{dobrinic2021examining}. 


On social platforms, one example of value misalignment between user and system is that many recommendation platforms seek to increase user engagement with content, even if that engagement is harmful to those who created or consume the content. One example of this misalignment is that \emph{``posts containing hate speech may garner many reactions and comments, but that does not mean that a person needs or wants to see hate speech at the top of their feed,''} \cite{devito2021values}. 

Finally, the common objective in recommendation to keep users scrolling, viewing, or engaging with content has sometimes led to very adverse effects for users. In the case of TikTok, one study showed that users felt so understood by its personalized algorithms that they reported feeling \emph{addicted} to the platform \cite{siles2021most}. In other studies, social media addiction has proven to be statistically correlated with mental health challenges, lower self-esteem, and reduced academic performance \cite{hou2019social, lin2016association}. In some cases, users who were already experiencing poor mental health tried to use social media to improve their mood, and \emph{``when this need is not met, their mental condition tends to become worse... Thus, the relation between poor mental health and social media addiction is likely to be bidirectional,''} \cite{hou2019social}. In these scenarios, objective misalignment between the user (who's goal was to improve their mental health) and system (who's goal was to keep the user scrolling) led to more negative consequences than just a dislike for the algorithm -- it unintentionally caused users to worsen their already poor mental health status.

\subsection{Lack of User Agency and Control}
\label{agency}
Another cause for algorithmic hate is when users feel a lack of ability to ``tame their algorithm,'' \cite{simpson2022tame}. When users feel misunderstood, harmed by, addicted to, or dissatisfied with personalization, they often seek to control the algorithm through actions of their own. When these actions do not result in the outcomes users might have expected from their personal folk theories of the system, some users have sought to gain their agency back through activism or protest. Protests on Instagram and Twitter included viral use of the \#RIPInstagram or \#RIPTwitter hashtags in an attempt to signal dissatisfaction to the platform about personalized feeds \cite{skrubbeltrang2017ripinstagram, riptwitter}. On TikTok, some LGBTQIA+ users resorted to curating their own `For You Page' through meticulous attention to which videos they allowed themselves to finish on the platform, fearing that watching the `wrong' video might lead TikTok to label them incorrectly and distort their carefully curated digital identity on the platform \cite{simpson2022tame}. In other scenarios, TikTok users have taken collective action to resist the algorithm's suppression of content from BIPOC creators \cite{karizat2021algorithmic}. Algorithmic resistance is not confined to the digital world: in the case of the A-level algorithm scandal, thousands of people protested in the streets of the United Kingdom carrying signs that read, ``\#FuckTheAlgorithm,'' \cite{benjamin2022fuckthealgorithm}.

Though independent auditing and collective action have helped users gain a sense of control and agency over personalized algorithms, this approach can sometimes \emph{increase} algorithmic hate rather than alleviate it. In the following section, we explore actions that can be taken by the \emph{engineers} who create the algorithms instead of end-users in an effort to decrease algorithmic hate and improve the relationship between recommender systems and their users.

\section{What Alleviates Algorithmic Hate?}
\label{alleviates}
Algorithmic hate, aversion, dislike, or irritation negatively impacts the relationship between user and system and can result in adverse side effects such as lowered user retention, low user satisfaction, or algorithmic harm. We claim that the primary causes of algorithmic hate are the unintended results of decisions made while designing these systems. Thus, if we can confront and intervene on the technological or design approaches that lead to algorithmic hate, we can help alleviate this concern in the future. In this section, we detail several promising directions in the RecSys research community that could help in this endeavor. 

\subsection{Human-Centered Recommender System Design}
In Section \ref{misalignment}, we described scenarios where a mismatch in objectives between stakeholder(s) and system resulted in algorithmic hate. Aligning values between and across all stakeholder groups can be difficult in complex systems. Fortunately, the field of human-centered machine learning confronts this challenge. \emph{``The goal of human-centered recommender systems research is to design the algorithms and interactions of recommender systems to better fulfill the goals of users and of the organizations engaging with these users,''} \cite{konstan2021human}. In human-centered machine learning, one goal is to help create systems that understand users better, while also helping users understand the system better \cite{riedl2019human}. The former seeks to improve issues of representative harm, while the latter seeks to improve user knowledge of the system -- both of which are causes of algorithmic hate identified in this paper. One key component to human-centered work in recommendation is working with real people through user studies. Though user studies were once a foundational aspect of RecSys research, monetary and time constraints combined with a newfound ease in offline experimentation has led the discipline, especially academic practitioners, away from working with real users and towards simulating user behavior and preferences instead. In order to truly confront some of the causes of algorithmic hate through human-centered methods, studies with real users will be an essential component.

\subsection{Transparency}
One theme relevant to algorithmic hate is users' lack of understanding of the systems they use. If users do not understand an algorithm, they cannot reliably control how their actions impact their personalized content -- leading to dissatisfaction, harm, or even protest. One method to combat this is to educate users about how the algorithm works through transparency in design. Previous work in the RecSys discipline has explored methods for explaining why users are being recommended content \cite{tintarev2007survey}, and which types of explanations are most helpful for improving user trust and adoption of the system \cite{woodruff2018qualitative, sonboli2021fairness}. However, simply telling users about how the algorithm works will not always reduce algorithmic hate. When users are told how the algorithm works and they fundamentally \emph{dislike} the design of the algorithm, this can lead to algorithmic aversion -- when users choose to not use algorithms after discovering its imperfections \cite{dietvorst2015algorithm}. Thus, future RecSys research could equally benefit from more human-centered approaches to explore new methods for designing recommender systems in a way that match users' understanding and expectations. If systems are designed to better fit user needs, then transparency is more likely to improve user trust rather than degrade it.


\subsection{Increased User Control \& Agency}
Another common theme in cases of algorithmic hate is a feeling that users lack control or agency over ``the algorithm''. While transparency and education about how the system works is an important first step in improving agency, additional steps could include increased configurability of recommender systems \cite{felfernig2013toward}. Configurability is not a simple solution as it requires presenting users with enough information to take action without causing information overload. Future work in this domain will require striking a balance between predicting what users prefer and allowing users to select their own preferences \cite{falkner2011recommendation, garcia2011information}. Some researchers have called for greater user agency through a decoupling of social media platforms from algorithmic curation \cite{fukuyama2021save}. This so-called ``middleware'' solution proposes that users be able to choose from a marketplace of front-end algorithms, with different providers competing to meet different user needs for curation. Social media platforms would not voluntarily cede this much control over their user base, however, so it seems unlikely that this type of solution could be achieved without regulatory action.

\section{Conclusions and Future Work}
Though personalization algorithms were originally designed to improve user experience on digital platforms, over time they have led to undesired negative consequences such as algorithmic harm and user dissatisfaction. When users feel harmed or dissatisfied by personalized systems, this can lead to algorithmic hate -- negative behaviors and attitudes felt from users towards an algorithmic system. In this work, we outlined various causes for algorithmic hate in recommendation, such as lack of knowledge about the system, representational or allocative harms, objective mismatches between user and system, and a lack of user agency or control. Since algorithmic hate has negative consequences for both users and the system, we propose several promising directions for future RecSys research to help combat algorithmic hate by design. We hope that with increased attention to human-centered approaches, we can help improve transparency and user agency in personalized recommendations -- with the goal to cultivate more positive relationships between recommender systems and their users in the future.

\begin{acks}
This research was supported by the National Science Foundation under grant IIS-2107577.
\end{acks}

\bibliographystyle{ACM-Reference-Format}
\bibliography{refs}

\end{document}